# Probing Aqueous Electrolytes with Fourier Spectrum Pulse-Echo Technique

Barnana Pal[1] and Sudakshina Roy[2]

Saha Institute of nuclear Physics,
Condensed Matter Physics Division,
1/AF, Bidhannagar, Kolkata-700064, India.
e-mail: barnana.pal@saha.ac.in

Abstract:

The nature of variations of ultrasonic wave velocity($v$) and attenuation constant($\alpha$) with the concentration(c) in aqueous solutions of NaCl, KCl and CsCl are investigated at room temperature($25^0$C) at 1MHz and 2 MHz wave frequency. Fourier Spectrum Pulse-Echo (FSPE) technique is used to achieve better accuracy in measurement, particularly for $\alpha$ measurement. In NaCl and KCl abrupt changes in the values of $v$ and $\alpha$ are noticed at particular solution concentrations whereas, for CsCl almost smooth variation in $v$ is observed over the whole concentration range for both of the frequencies 1 and 2 MHz . The nature of variation in $v$ and $\alpha$ in these solutions are analyzed in view of other spectroscopic studies. The well-known Jones-Dole equation that explains the viscosity($\eta$) variation in many electrolyte solutions, offers satisfactory fits to the experimental velocity variation with parameter values characteristics of the sample.

Keywords: Fourier spectrum, pulse echo method, ultrasonic propagation constants, aqueous electrolytes, Jones-Dole equation.

## 1. Introduction:

Characterization of aqueous electrolytic solutions by various experimental and simulation techniques reveal interesting features of the structure and dynamics of these solutions. Depending on the concentration of the solution, the ions present play significant role on "structure making" or "structure breaking" of liquid water which is believed to form an ordered network structure through hydrogen bonds [1-3] in the absence of any impurity material. Addition of electrolytes in water cause dissociation of the salt into ions and consequent rearrangement of ions and water molecules forming hydration shells. Such description for the structure of water and aqueous electrolytes is accepted in general but the question arises in the issue of the quantitative description and corresponding experimental verification. Numerous experimental techniques have been used to study the structure and dynamics of water and aqueous electrolytes. Some of these are X-ray diffraction [2-6], X-ray absorption spectroscopy[7,8]; Neutron diffraction [2,9-11], Raman spectroscopy [6,12-14], NMR[15-17], IR spectroscopy [18], Femtosecond elastic second harmonic scattering [19]; Dielectric relaxation spectroscopy [20-22], Ultrasonic spectroscopy [23,24], Dynamic light scattering (DLS) [25] and Molecular dynamics simulation [5,26-29]. These studies help in understanding the system over an extended length scale starting from short range atomic level to long range bulk properties. It is accepted that ion-dipole interaction is an important factor causing perturbation to the complex network structure of pure water formed through hydrogen bonds. The strong electric field around an ion forces the dipolar water molecules to rearrange themselves in hydration shells. Authentic information regarding the spatial extent of these hydration shells, the ionic interaction strength beyond the hydration shells and the role of solution concentration in monitoring structural effects are still lacking. The present report describes the experimental observations on the ultrasonic propagation properties, viz., the velocity ($v$) of propagation and attenuation constant ($\alpha$), in aqueous solutions of NaCl, KCl and CsCl over a wide concentration (c) range up to near saturation. Fourier Spectrum Pulse-Echo (FSPE) method [30] is used to determine the parameters. The method is described in sec.2, results are discussed in sec.3 and sec.4 gives summary and conclusion.

---

[1] Corresponding author
[2] CSIR-Central Glass and Ceramic Research Institute, Kolkata (retired), e-mail: chatterjeesudakshina@gmail.com

## 2. The Experiment:

Fourier spectrum pulse-echo (FSPE) method is used for the determination of ultrasonic velocity $v$ and attenuation constant α in the electrolyte solutions. The detailed description of FSPE is available in ref. [30]. The pulsed electrical signal is generated using Ultran HE900 High Energy RF burst generator. RF bursts of 1 MHz and 2 MHz centre frequency with repetition time ~10 ms and p-p voltage ~200 V are used as the input to the transducer. The echo train is captured with YOKOGAWA DL 1640 200MS/s 200MHz digital oscilloscope. Data acquisition in the average mode reduces the noise level. Data acquisition time is kept large such that the complete echo signal starting from the first echo to the smallest visible echo is contained well within the time span. This is to ensure finer details of the frequency components in the Fourier spectrum. The frequency spectrum shows amplitude peaks at certain frequencies f which are perfectly equidistant in the frequency scale with spacing

$$\Delta f = v/2l \qquad (2.1)$$

l being the length of the sample. The attenuation constant α is related to the Q-value of the resonance peak at frequency f by the equation

$$\alpha = \pi f/Q \qquad (2.2)$$

Under favorable experimental condition, the peak frequency components are found equidistant with high degree of accuracy. From the FFT peaks $v$ and α are determined using relations (2.1) and (2.2). These are found to be more consistent and accurate. The estimated error in measuring $v$ is less than 0.5% and that for α is less than 20%.

The experiment is done at room temperature maintained at $25^0$ ($\pm 1^0$) C. The salt solution is taken in a cylindrical glass container. A perfectly plain glass plate is placed at the bottom of the container which acts as the reflector for the ultrasound signal. The transducer is mounted keeping its face parallel to the reflector surface on one end of a horizontal rod. The other end of the rod is attached to a vernier scale such that the transducer can move vertically upwards or downwards and the length l of liquid column can be measured by the vernier scale with an accuracy of 0.002 cm. Aqueous solutions of NaCl, KCl and CsCl of different molar concentrations c are prepared using Elix ultrapure de-ionized water as the solvent and 99.90% pure salts from Loba Chemie Pvt. Ltd. as the solute. The estimated error in concentration is less than 1%.

## 3. Experimental Results:

The concentration dependence of $v$ and α measured over the region starting from a low value to near saturation are described for the three solutions separately in the following three sections.

### 3.1 Measurement with NaCl solution:

Figure1(a) and (b) show the variation of $v$ with c for aqueous NaCl solution at ultrasonic frequencies 1MHz and 2 MHz respectively. An overall increase in $v$ is observed with the increase in c for both of the frequencies and around c = 3.0M an abrupt change is noticed which is more prominent at 2 MHz wave frequency. Figure2(a) and (b) show the change of α with c at 1 MHz and 2 MHz respectively. With increasing c, α also increase for the two frequencies showing abrupt increase near c = 2.0M and 4.0M although the change in α near c = 0.2M is not very prominent at 1 MHz.

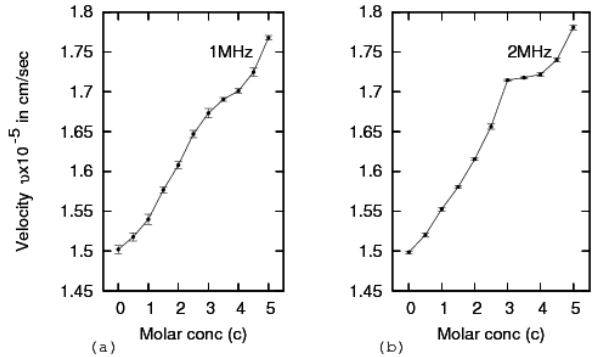

Figure1: Velocity $v \times 10^{-5}$ in cm/sec as function of molar concentration c for aqueous NaCl at (a) 1MHz and (b) 2MHz.

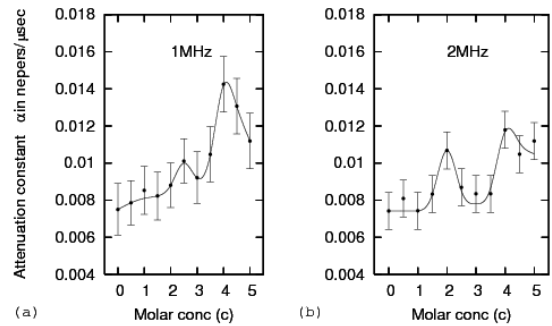

Figure2: Attenuation constant α in nepers/μsec vs. molar concentration c for aqueous NaCl at (a) 1MHz and (b) 2MHz.

### 3.2 Measurement with KCl solution:

For KCl solution, variation of $v$ as a function of c is shown in figures 3(a) and (b) for 1 MHz and 2 MHz respectively. In this case also an abrupt change in velocity value is observed which is more prominent at higher frequency, i.e., at 2 MHz. Variation of α with c is shown in figures 4(a) and (b) for 1 MHz and 2 MHz respectively. In this case α shows more or less smooth increase with increasing c with a peak appearing near 2.5 M at 1 MHz and for 2 MHz, abrupt rise in α value is observed at two concentrations near 1M and 2.5M.

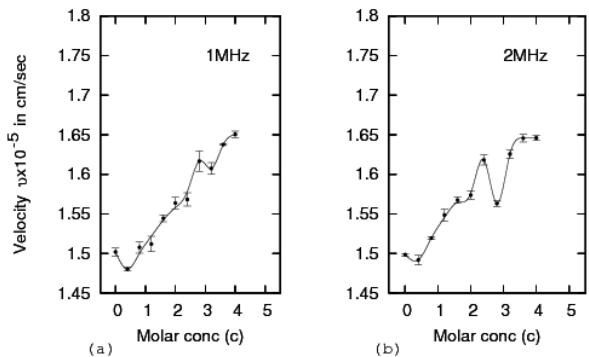

Figure3: Velocity $v \times 10^{-5}$ in cm/sec as function of molar concentration c for aqueous KCl at (a) 1MHz and (b) 2MHz.

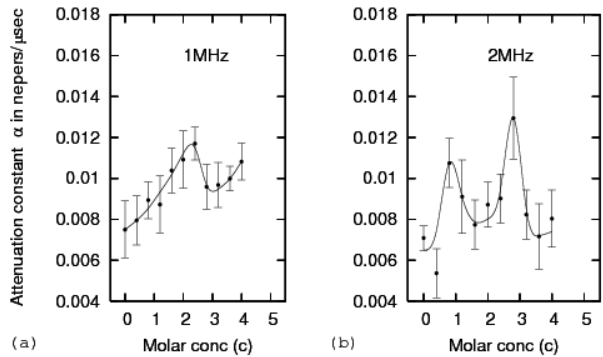

Figure4: Attenuation constant α in nepers/μsec vs. molar concentration c for aqueous KCl at (a) 1MHz and (b) 2MHz.

### 3.3 Measurement with CsCl solution:

For aqueous CsCl, the variation of $v$ with c is given in Figures 5(a) and (b) for 1MHz and 2MHz respectively. It is seen that $v$ decreases with the increase of c at lower concentration, attains a minimum and again starts increasing. Plot of α vs. c at 1 MHz and 2 MHz is shown in figures 6(a) and (b) respectively. At 1 MHz, α remains almost constant up to c = 5M and then starts increasing a little. For 2 MHz, an overall increase in α is noticed with the indication of abrupt rise at c = 2M and 6M.

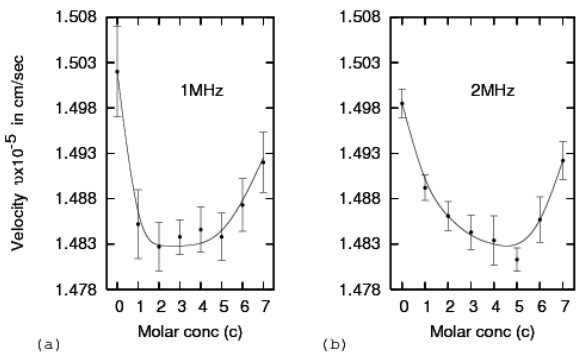
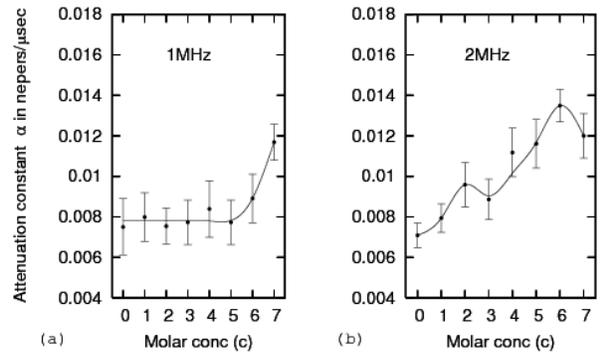

Figure5: Velocity $v \times 10^{-5}$ in cm/sec as function of molar concentration c for aqueous CsCl at (a) 1MHz and (b) 2MHz.

Figure6: Attenuation constant $\alpha$ in nepers/μsec vs. molar concentration c for aqueous CsCl at (a) 1MHz and (b) 2MHz.

## 4. Discussion:

A comparison of figures 1, 3 and 5 shows that for NaCl and KCl, $v$ shows an overall increase with the increase in c, and for CsCl, $v$ decreases with the increase in c, attains a flat minimum and then increases a little. The change in $v$ for CsCl is much smaller compared to the change observed in NaCl and KCl. For KCl, $v$ shows a small but noticeable decrease in c at low c. This observation is consistent with the viscosity data available in the literature. Figure 7 illustrates typical nature of variation of viscosity η with c, (a) for NaCl [31] and (b) for KCl and CsCl [32]. For NaCl, η increases smoothly with increasing c and similar behaviour is noticed for variation of $v$ with c (fig.1) at both of the two frequencies. In addition, there is indication of an anomalous rise near c = 3 M and this is more prominent at 2MHz. Fig 7(b) shows a decrease in η at lower c for KCl and CsCl. For KCl, η decreases a little up to c = 1 M and then it starts increasing smoothly. For the present study with KCl, a similar decrease in $v$ is noticed at c ≤ 1 M for both of the two frequencies and then $v$ starts increasing. Anomalous change in $v$ is observed near c = 3 M. For CsCl, η decreases appreciably up to c= 2 M and then it increases slowly as is seen in fig.7(b). The variation of $v$ with c for CsCl follow similar nature, as seen in fig 5, for both of the frequencies 1 and 2 MHz. For CsCl, the variation is more or less smooth.

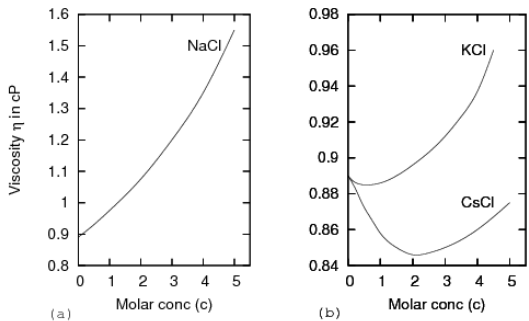

Figure7: Viscosity η in cP vs. molar concentration c for (a) NaCl, (b) KCl and CsCl.

For dilute solutions, c < 1M, the concentration dependence of viscosity in aqueous electrolytes is well represented by Jones-Dole semi-empirical equation [33-36],

$$\eta_r = \frac{\eta}{\eta_0} = 1 + Ac^{1/2} + Bc \quad \text{..................................(4.1)}$$

Here $\eta_r$ is the relative viscosity, η and $\eta_o$ are the viscosities of the solution and that of pure water respectively. The coefficients A, and B are constants dependent on the nature of the solute and solvent. Further these are functions of temperature and pressure. The co-efficient A is determined by ion-ion interactions and may be calculated theoretically [37]. The constant B, called the Jones-Dole coefficient, represents ion-solvent interactions. This B coefficient can be divided into ionic contribution. A positive B coefficient implies "structure making" whereas negative B coefficient indicates "structure Breaking". This basic equation has been further extended with inclusion of higher order terms in c for a satisfactory explanation for the viscosity results at higher concentrations [38,39].

$$\eta_r = \frac{\eta}{\eta_0} = 1 + Ac^{1/2} + Bc + Dc^2 \quad \ldots\ldots\ldots\ldots\ldots\ldots(4.2)$$

The coefficient D represents contribution from solute-solute interaction. Equation (4.2) is widely used for explaining viscosity data in many electrolyte solutions. Many attempts have been made to develop the theories and equations to establish an unique relation between viscosity $\eta_r$ and concentration c, applicable to the whole concentration region, not only for a single salt solution but also for mixed electrolyte solutions[40-45]. In this endeavour, it is found that an empirical relation with the inclusion of one additional term, $Ec^{3.5}$, is reasonably good in explaining viscosity data of many electrolytes over the full concentration range [46].

$$\eta_r = \frac{\eta}{\eta_0} = 1 + Ac^{1/2} + Bc + Dc^2 + Ec^{3.5} \quad \ldots\ldots\ldots\ldots\ldots\ldots(4.3)$$

In the present work a similar empirical relation is used to explain velocity data and to distinguish it from the viscosity equation we write the coefficients in small letters.

$$v_r = \frac{v}{v_0} = 1 + ac^{1/2} + bc + dc^2 + ec^{3.5} \quad \ldots\ldots\ldots\ldots\ldots\ldots(4.4)$$

Here $v_r$ is the relative ultrasound velocity, $v$ and $v_o$ are the velocities in the solution and in pure water respectively. The best fit coefficients for our experimental data for NaCl, KCl and CsCl are presented in Table-I. The viscosity B co-efficient and ionic radius[44] of $Na^+$, $K^+$, $Cs^+$, and $Cl^-$ ion are presented in Table-II. Figure8(a) and (b) show change in relative velocity $v_r$ with c along with Jones-Dole fit (solid line) in NaCl solution for 1 MHz and 2 MHz respectively. It is seen that the fit is good up to c ≈ 3.0 M

**Table-I: Best fit coefficients for $v_r$ vs. c plot with Jones-Dole type equation (4.4).**

| Salt | 1 MHz | | | | 2 MHz | | | |
|---|---|---|---|---|---|---|---|---|
| | a | b | d | e | a | b | d | e |
| NaCl | 0.0058 | 0.0126 | 0.0091 | - | 0.0056 | 0.0190 | 0.0084 | - |
| KCl | -0.1098 | 0.1536 | -0.0398 | 0.0038 | -0.1411 | 0.2404 | -0.0826 | 0.0092 |
| CsCl | -0.0212 | 0.0107 | -0.0010 | $2.5 \times 10^{-5}$ | -0.0101 | 0.0047 | -0.0008 | $3.5 \times 10^{-5}$ |

* rms of residues for standard fit is less than 0.006

**Table-II: Jones-Dole viscosity B-coefficient and ionic radius[44] for $Na^+$, $K^+$, $Cs^+$ and $Cl^-$ ion.**

| Ion | B in L.Mol$^{-1}$ | Ionic radius in Å |
|---|---|---|
| $Na^+$ | 0.085 | 1.02 |
| $K^+$ | -0.009 | 1.38 |
| $Cs^+$ | -0.047 | 1.70 |
| $Cl^-$ | -0.005 | 1.81 |

without the last term in eqn. (4.4). The viscosity B co-efficient for $Na^+$ ion is positive implying that it has a "structure making" effect in the aqueous solution. On the contrary, for $Cl^-$ ions B is negative implying "structure breaking" effect, but, the structure breaking effect is weaker compared to the structure making effect. Also, the ionic radius of an ion is considered to be one of the controlling factors by some researchers [10,47]. The ionic radius of $Na^+$ ion is small compared to that of $Cl^-$ ion in the solution. Smaller ionic radius means larger ionic charge density and thereby causing stronger ordering effect on neighboring water molecules. Thus the $Na^+$ ions produced due to the addition of NaCl in water boost up the structure making effect strengthening ion-water bonding and this is reflected as an increase in the ultrasonic velocity. A discontinuous change is observed above 3M which is more prominent at 2 MHz frequency. This may be due to the effect of $Cl^-$ ions causing a rearrangement in the ion-water structure. The variation of $v_r$ at higher c values cannot be fitted with Jones-Dole type equation (4.4) even with the inclusion of the term $ec^{3.5}$.

Figure 9(a) and (b) show changes in relative velocity $v_r$ with c along with Jones-Dole fit (solid line) in KCl solution for 1MHz and 2MHz respectively. Here the fit is reasonably good for $c \lesssim 3.0M$. For aqueous KCl solution, the viscosity B coefficients have small -ve values for both $K^+$ and $Cl^-$ ions indicating structure breaking effect in pure water and thereby causing an initial decrease in the viscosity η up to 1 M (fig 7(b)). The initial decrease in the relative velocity $v_r$ may also be attributed to this type of structure breaking effect. Above 1M an increase in η as well as $v_r$ is observed and this may be due to the formation of ion-water cluster induced by the ionic charge distribution of $K^+$ ions. For KCl also a discontinuity is observed near c = 3M which is more prominent at 2MHz.

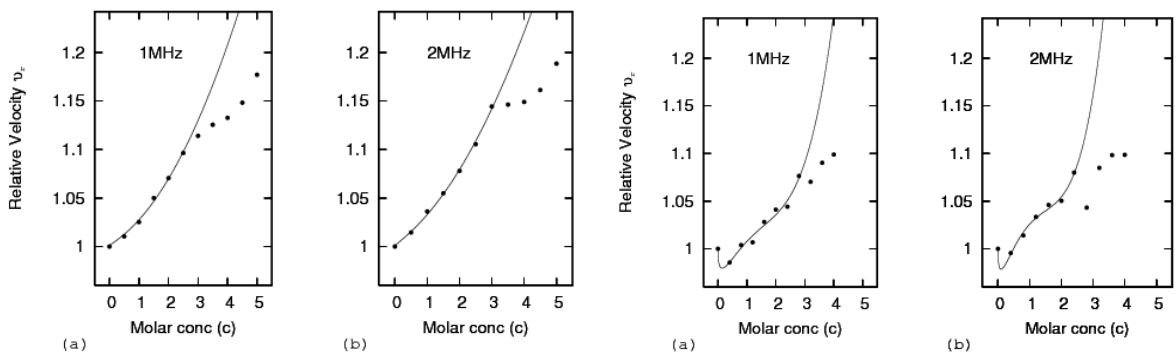

Figure8: Relative Velocity $v_r$ as function of molar concentration c for aqueous NaCl at (a) 1 MHz and (b) 2 MHz. Dots represent experimental points and the solid line represents Jones-Dole fit.

Figure9: Relative Velocity $v_r$ as function of molar concentration c for aqueous KCl at (a) 1 MHz and (b) 2 MHz. Dots represent experimental points and the solid line represents Jones-Dole fit.

For CsCl, the changes in relative velocity $v_r$ with c along with Jones-Dole fit (solid line) is given in figure 10(a) and (b) for 1MHz and 2MHz respectively. Here a reasonably good fit is obtained over the whole concentration range. The viscosity B coefficient for $Cs^+$ ion has a large -ve value indicating strong structure breaking effect. In aqueous CsCl solution both $Cs^+$ and $Cl^-$ act as structure breaking element in water as observed in KCl solution, but the effect is more prominent due to the presence of $Cs^+$ ions. The result is a considerable decrease in η (fig7(b)) as well as in $v_r$ with the addition of CsCl in water. At 1 MHz, $v_r$ remains almost unchanged from 1 M to 5 M and then increases a little. At 2 MHz, the decrease of $v_r$ up to 5 M is somewhat slower compared to that observed at 1 MHz. For CsCl no prominent discontinuity is observed.

Ionic radii of $Na^+$, $K^+$, $Cs^+$ and $Cl^-$ ions are in increasing order with charge densities in decreasing order and consequently contributing decreasing effect on water structure making. A comparative featureless structure around $K^+$ ions was reported by Sacco et. al. [45]. They identified CsCl as a strong structure breaker. Their observations are in agreement with the present study on the nature of variation of $v_r$ with c for the three salts NaCl, KCl and CsCl in aqueous solutions. The curves showing ultrasonic velocity $v_r$ in NaCl for both of the frequencies lie over the corresponding curves for KCl solution for the entire concentration range indicating stronger structure making ability of $Na^+$ ions over $K^+$ ions. In contrast, the nature of same plots obtained for CsCl solution strengthens the idea of structure breaking property of $Cs^+$ ions. At higher c values, the contribution from the $Cl^-$ ions play some role as structure maker influencing the orientation of the nearby water molecules and this becomes responsible for the positive slope of the $v_r$ vs. c plot.

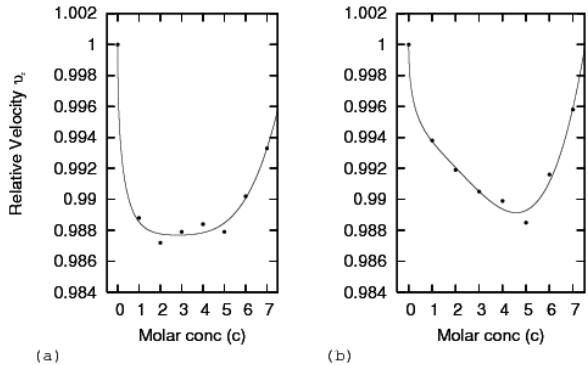

Figure 10: Relative velocity $v_r$ as function of molar concentration c for aqueous CsCl at (a) 1 MHz and (b) 2 MHz. Dots represent experimental points and the solid line represents Jones Dole-fit

Regarding attenuation constant α, we see from figures 2, 4 and 6 that, besides an overall increase in α with c, there are sharp rises in α at particular values of c and this is more prominent at 2 MHz (figs 2(b), 4(b) & 6(b)). For NaCl the peaks are near 2M and 4M (fig 2(b)) and these may be associated with a change in slope in corresponding $v$ vs. c plot (fig 1(b)) near 3M and 4M. For KCl α shows sharp rise near 1M and 3M (fig 4(b)) and corresponding $v$ vs. c plot (fig 3(b)) shows discontinuity near 0.5 M and 3M. For CsCl (fig 6(b)) α peaks appear at 2M and 6M though the peak at 2M is not so prominent. Corresponding $v$ vs. c plot show change in slope near 2M and 5M (fig 5(b)). The change in slope or discontinuity in $v$ vs. c plot and associated attenuation peak is an indication of structural transitions in the aqueous solutions. The difference in structure making and structure breaking effects of the ions in NaCl or the difference in structure breaking effects of the ions in KCl as indicated by viscosity B coefficients and ionic radii, play significant role in the formation of ion-water clusters replacing prevailing network structure in pure water. For CsCl, since $Cs^+$ ion possesses strong structure breaking effect, these ions play significant role in breaking the network structure in pure water leaving only a bare possibility of ion-water cluster formation at high concentrations only.

## 5. Conclusion:

In the present study Fourier Transform Pulse-echo (FTPE) method is used for the measurement of ultrasonic propagation parameters $v$ and α in aqueous electrolyte solutions of NaCl, KCl and CsCl. The variation of these parameters with solution concentration c is determined over the full concentration range starting from a low value up to near saturation at room temperature 25°C. Using FTPE we have been able to measure $v$ with 0.5% accuracy and α within 20% accuracy. The interesting point is the observation of sharp rise in α at specific c values which are associated with change in slope or discontinuity in the respective $v$ vs. c plot. The effect is more prominent at higher ultrasonic frequency, i.e. at 2MHz for NaCl and KCl. The results indicate changes in the pure water network structure and subsequent formation of ion-water cluster mediated by the presence of the ions in the solution. A phenomenological analysis of the observed effect is presented considering Jones-Dole viscosity B coefficients and ionic radii for the ions. A similar Jones-Dole type equation is used to fit the variation of $v$ with c and it is observed that such a fit is possible only in the lower concentration region for NaCl and KCl. For CsCl however, the fit is reasonably good over the full concentration range.

**Acknowledgement:** The authors are grateful to Sankari Chakraborty and Papia Mondal for their assistance and technical help in conducting the experiment.

**References:**

[1] Marcus Y, Effect of ions on the structure of water: structure making and breaking, Chem Rev 109 (2009) 1346-1370.

[2] Ohtaki H, Structure and dynamics of hydrated ions, Chem Rev 93 (1993) 1157-1204.

[3] Ohtaki H, Fukushima N, A structural study of saturated aqueous solutions of some alkali halides by X-ray diffraction, J Sol Chem 21(1) (1992) 23-38.

[4] Vinogradov E V, Smirnov P R, Trostin V N, Structure of hydrated complexes formed by material ions of groups I-III of the Periodic Table in aqueous electrolyte solutions under ambient conditions, Russ Chem Bull 52 (2003) 1253-1271.

[5] Bouazizi S, Nasr S, Jaidane N, Bellissent-Funel M-C, Local order in aqueous NaCl solutions and pure water: X-ray scattering and Molecular Dynamics simulation study, J Phys Chem B 110 (2006)23515-23523.

[6] Li F, Yuan J, Li D, Li S, Han Z, Study on the structure of aqueous potassium chloride solutions using the X-ray diffraction and Raman spectroscopy methods, J Mol Struc 1081 (2015) 38-43.

[7] Waluyo I, Nordlund D, Bergmann U, Schlesinger D, Pettersson L G M, Nilsson A, A different view of structure-making and structure-breaking in alkali halide aqueous solutions through x-ray absorption spectroscopy. J. Chem. Phys. 140 (2014) 244506.

[8] Nilsson A, Nordlund D, Waluyo I, Huang N, Ogasawara H, Kaya S, Bergmann U, Näslund L-Å, Öström H, Wernet P, Andersson K J, Schiros T, Pettersson L G M, X-ray absorption spectroscopy and X-ray Raman scattering of water and ice; an experimental view, J. Electron Spectrosc. 177 (2010) 99–129.

[9] Novikov A G, Rodnikova M N, Savostin V V, Sobolev O V, The study of hydration effects in aqueous solutions of LiCl and CsCl by inelastic neutron scattering, J Mol Liq 82 (1999) 83-104.

[10] Mancinelli R, Botti A, Bruni F, Ricci M A, Sopar A K, Perturbation of water structure due to monovalent ions in solution, Phys Chem Chem Phys 9 (2007) 2959-2967; Hydration of sodium, potassium and chloride ions in solution and the concept of structure maker/breaker, J phys chem B 111 (2007) 13570-13577.

[11] Soper A K, Weckström K, Ion solvation and water structure in potassium halide aqueous solutions, Biophys Chem 124 (2006) 180-191.

[12] Sun Q, Zhao L, Li N, Liu J, Raman spectroscopic study for the determination Cl- concentration (molarity scale) in aqueous solutions: applications to fluid inclusions, Chem Geology 272 (2010) 55-61.

[13] Terpstra P, Combes D, Zwick A, Effect of salts on dynamics of water: a Raman spectroscopy study, J chem Phys 92(1) (1990) 65-70.

[14] Chumaevskii N A, Rodnikova M N, Sirotkin D A, Cationic effect in aqueous solutions of 1:1 electrolytes by Raman spectral data, J Mol Liq 91 (2001) 81-90.

[15] Engel G, Hertz H G, On the negative hydration. A nuclear magnrtic relaxation study, Ber Bunsen-ges Phys Chem 72 (1968) 808-834.

[16] Sacco A, Structure and dynamics of electrolyte solutions. A NMR relaxation approach, Chem Soc Rev 23 (1994) 129-136.

[17] Yoshida K, Ibuki K, Ueno M, Estimated ionic B-coefficients from NMR measurements in aqueous electrolyte solutions, J Sol Chem 25 (1996) 435–453.

[18] Zhang Q, Chen H, Wu T, Jin T, Pan Z, Zheng J, Gao Y, Zhuang W, Chem Sci 8 (2017) 1429-1435.

[19] Chen Y, Okur H I, Gomopoulos N, Macias-Romero C, Cremer P S, Poul B. Petersen P B, Tocci G, David M. Wilkins D M, Liang C, Ceriotti M, Roke S, Electrolytes induce long-range orientational order and free energy changes in the H-bond network of bulk water, Sci Adv 2 (2016) e1501891(1-8).

[20] Kaatze U, The dielectric properties of water in its different states of interaction, J Sol Chem 26 (1997) 1049-1112.

[21] Buchner R, Hefter G T, May P M, Dielectric relaxation of aqueous NaCl solutions. J. Phys. Chem. A 103 (1999) 1–9.

[22] Chen T, Hefter G, Buchner R, Dielectric spectroscopy of aqueous solutions of KCl and CsCl, J Phys Chem A 107 (2003) 4025-4031.

[23] Voleišienė B, Voleišis A, Ultrasound velocity measurements in liquid media, Ultragarsas(Ultrasound) 63 (2008) 7-19.

[24] Gucker F T, Chernick C L, Roy-Chowdhury P, A frequency-modulated ultrasonic interferrometer: Adiabatic compressibility of aqueous solutions of NaCl and KCl at $25^0$C, PNAS 55 (1966) 12-19

[25] Georgalis Y, Kierzek A M, Saenger W, Cluster formation in aqueous electrolyte solutions observed by dynamic light scattering, J Phys Chem 104 (2000) 3405-3406.

[26] Balbuena P B, Johnston K P, Rossky P J, Hyun J-K, Aqueous ion transport properties and water reorientation dynamics from ambient to supercritical conditions, J Phys Chem B 102 (1998) 3806-3814.

[27] Chowdhuri S, Chandra A, Molecular dynamics simulations of aqueous NaCl and KCl solutions: Effects of ion concentration on the single-particle, pair, and collective dynamical properties of ions and water molecules, J Chem Phys 115 (2001) 3732-3741.

[28] Guàrdia E, Laria D, Martí J, Hydrogen Bond Structure and Dynamics in Aqueous Electrolytes at Ambient and Supercritical Conditions, J Phys Chem B 110 (2006) 6332-6338.

[29] Liu Y, Lu H,Wu Y, Hu T, Li Q, Hydration and coordination of K+ solvation in water from *ab initio* molecular-dynamics simulation, J Chem Phys 132 (2010) 124503.

[30] Pal Barnana, Fourier Transform Pulse-Echo for Acoustic Characterization, arXiv:1801.04851v1.

[31] Ozbek H, Viscosity of aqueous sodium chloride solutions from 0-$150^0$C, Lawrence Berkeley National Laboratory, http://escholarship.org/uc/item/3ip6n2bf (2010).

[32] Goldsack D E, Franchetto R, The viscosity of concentrated electrolyte solutions. I. Concentration dependence at fixed temperature, Can J Chem 55 (1977)1062.

[33] Jones G, Dole M, The Viscosity of Aqueous Solutions of Strong Electrolytes with Special Reference to Barium Chlorides, J Am Chem Soc 51 (1929) 2950-2964.

[34] Jones G, Talley S K, The Viscosity of Aqueous Solutions as a Function of the Concentration. II. Potassium Bromide and Potassium Chloride, J Am Chem Soc 55 (1933) 4124-4125.

[35] Jones G, Talley S K, The Viscosity of Aqueous Solutions as a Function of the Concentration, J Am Chem Soc 55 (1933) 624-642.

[36] Stokes R H, Mills R, Viscosity of Electrolytes and Related Properties, Pergamon, New York, 1965.

[37] Falkenhagen H, Theorie der Elektrolyte; Hirzel: Leizig, 1971.

[38] Kaminsky M, Experimental study of the concentration and temperature dependence of the viscosity of aqueous solutions of strong electrolyte. I. Z Phys Chem 8 (1956) 173.

[39] Kaminsky M, Experimental Investigation on Viscosity of Dilute Aqueous Strong Electrolyte Solutions at Different Temperatures III. KCl, $K_2SO_4$, $MgCl_2$, $BeSO_4$ and $MgSO_4$, Z Phys Chem 12 (1957) 206-231.

[40] Goldsack D E, Franchetto A A, The viscosity of Concentrated Electrolyte Solutions III. A Mixture Law, Electrochim Acta 22 (1977) 1287-1294.

[41] Nowlan M F, Doan T H, Sangster J, Prediction of the viscosity of Mixed electrolyte Solutions from Single-Salt Data, Can J Chem Eng 58 (1980) 637-642.

[42] Out D J P, Los J M, Viscosity of Aqueous Solutions of Univalent Electrolytes from 5 to 95$^o$C, J Sol Chem, 9(1) (1980) 19-35.

[43] Horvath A C, Handbook of aqueous electrolyte solutions, Ellis Horwood Limited, London, 1985.

[44] Marcus Y, Viscosity B-Coefficients, Structural Entropies and Heat Capacities, and the Effects of Ions on the Structure of Water, J Sol Chem 23 (1994) 831-848.

[45] Sacco A, Weingartnert H, Braun B M, Holz M, Study of the Structure-breaking Effect in Aqueous CsCl Solutions Based on H20/D,0 Isotope Effects on Transport Coefficients and Microdynamical Properties, J Chem Soc Faraday Trans, 90(6) (1994) 849-853.

[46] Hai-Lang Z, Shi-Jun H, Viscosity and Density of Water + Sodium Chloride + Potassium Chloride Solutions at 298.15 K, J Chem Engg Data 41 (1996) 516-520.

[47] Hribar B, Southall N T, Vlachy V, Dill K A, How Ions Affect the Structure of Water, J Am Chem Soc, 124 (2002) 12302.